# How can heat maps of indexing vocabularies be utilized for information seeking purposes?

Peter Mutschke, Karima Haddou ou Moussa

GESIS – Leibniz Institute for the Social Sciences, Cologne, Germany
{peter.mutschke,karima.haddououmoussa}@gesis.org

**Abstract.** The ability to browse an information space in a structured way by exploiting similarities and dissimilarities between information objects is crucial for knowledge discovery. Knowledge maps use visualizations to gain insights into the structure of large-scale information spaces, but are still far away from being applicable for searching. The paper proposes a use case for enhancing search term recommendations by heat map visualizations of co-word relationships taken from indexing vocabulary. By contrasting areas of different "heat" the user is enabled to indicate mainstream areas of the field in question more easily.

## 1  Introduction

During the past decade interactive search interfaces [1] have emerged as important in Information Retrieval (IR) research. The insight that the success of information seeking mainly depends on the ability of an information system to properly support interaction between user and system has led to the establishment of Interactive Information Retrieval (IIR) as a specific discipline within IR research. Corresponding to this, whole-session retrieval issues [2] as well as search interfaces [3,4,5] became a focal point in research, in particular with respect to exploratory searching [4,6]. Due to the enormous increase of information spaces bibliometric enhanced IR models addressing non-textual attributes of the domain under study became also more and more important at the same time. This is particular true for the case of scholarly searching [7,8]. Moreover, studies in interactive information seeking behavior have confirmed that the ability to browse an information space in a structured way by exploiting similarities and dissimilarities between information objects is crucial for knowledge discovery [9,10].

Knowledge maps, on the other hand, use visualizations to gain insights into the structure of large-scale information spaces. They can take very different forms such as network visualizations, heat maps, tree maps or geographic map like arrangements of information spaces [11-17]. However, knowledge maps are still far away from being applicable as search interfaces for Digital Libraries. Most maps are static visualiza-tions made for special purposes, and hence neither interactive [18] nor dynamic, i.e. they do not adapt to the change of user perspective to an information space during interaction. Thus, combining knowledge mapping with IR is still a challenging research issue. There are just a few examples where visual concepts also used in knowledge mapping have been applied to information systems, such as DANSEasy[1], which visualizes an archive's category structure and its content in form of a dendrogram and a tree map to be used as a navigation tool through the information space. Another example is PepBank[2] which uses heat maps for visualizing and refining search results.

Interface studies have shown that a simple spatial interface layout performs better than complex ones [3]. Heat maps are simple visualizations of data in a color-coded 2-dimensional matrix where cells have a particular color index indicating remarkable values of the matrix. To the best of our knowledge, heat maps have not been used for the query formulation process so far. This position paper discusses a use case for using heat map visualizations of relationships between indexing terms to enhance search term recommendations.

## 2 Using Heat Maps of Co-Word Relationships for Searching

A particular point of failure of current information systems is the vagueness between user search terms and the terms used for indexing the documents to be retrieved, i.e. the *indexing terms* which are usually based on a controlled vocabulary such as a thesaurus [19]. Search term recommenders (STRs) provide models that map a user's search term to more appropriate terms. The model proposed by [7], for instance, maps search terms to indexing terms on the basis of a co-word analysis and recommends indexing terms that strongly co-occur with the search term. The expectation here is that retrieval quality will increase when indexing terms are used for searching. Indeed, a retrieval evaluation showed that the use of STRs relying on controlled vocabulary has a great potential to improve the precision of a search significantly [7].

However, search term recommendations usually appear in forms which do not assist the user in locating the information need within the wider information space. The STR provided by the Social Science literature portal sowiport[3], for instance, displays recommended indexing terms in a drill down menu (see Fig. 1, cp. [7]). From that list the user can select more appropriate terms for searching. The major problem however is that the user's choice is not facilitated by further potentially helpful information, in particular structural information about the semantic contexts in which a recommended term appears.

---

[1] http://www.drasticdata.nl/ProjectDANSEasy/indexMultipleAssignments.htm
[2] http://pepbank.mgh.harvard.edu/
[3] http://sowiport.gesis.org/

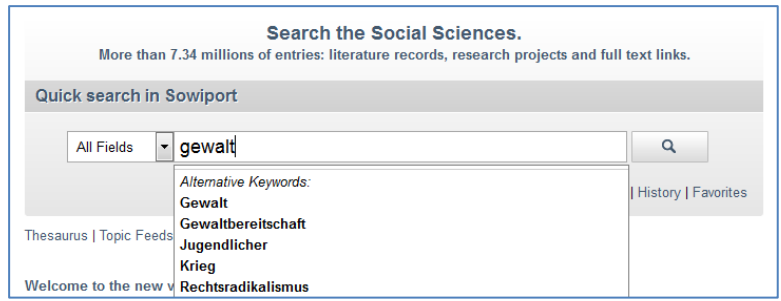

**Fig. 1.** Search Term Recommendations in Sowiport. Recommended indexing terms strongly co-occurring with the search term are displayed in a one-dimensional drilldown menu[4].

Our approach therefore is to extend the initial list of term suggestions by further indexing terms that frequently co-occur with the search term as well as with the initially suggested terms. This yields a two-dimensional space of co-word relationships where the initially recommended terms represent one dimension (first order terms) and the indexing terms co-occurring with the search term as well as with first order terms represent the other dimension (second order terms). A natural way to visualize such a matrix of co-word relationships is a heat map visualization (see Fig. 2) which displays the strength of each relationship (its "heat") by a color on a scale ranging from red (indicating high values) to blue (indicating low values). To keep it simple just the frequency of co-occurrences was taken to indicate the strength of a co-word relationship (see numbers in the cells of Figure 2[5]). The background color code of the cells of the map is calculated according to the ratio of the individual frequencies to the maximum and minimum values of the matrix. By this color-coded visualization the heat map is divided into "hot" (red), "warm" (yellow, green) and "cold" (blue) areas. Thus, red cells point to topic combinations which appear most frequently, blue cells correspond to areas which appear least frequently in the map. Red cells therefore represent issues which are more heavily discussed in the research field ("hot" topics of the fields). Accordingly, blue cells represent issues which are less heavily discussed (compared to "hot" fields). Hence, for red areas the user can expect to find more documents than for blue areas.

---

[4] Here, the search term "violence" is mapped to the indexing terms "Gewalt" ("Violence"), "Gewaltbereitschaft" ("Propensity to violence"), "Jugendlicher" ("Adolescent"), "Krieg" ("War"), and "Rechtsradikalismus" ("Right-wing radicalism"), based on the thesaurus for the Social Sciences provided by GESIS.

[5] The map displays the frequency of documents containing the respective term combination (see numbers in the cells as well as numbers next to the terms). Thus, the total number of documents containing the search term "Violence" is 9718; the total number of documents containing the search term "Violence" and the first order term "Adolescent" is 1934; the total number of documents containing the search term "Violence" and the second order term "Right-wing radicalism" is 846; and the total number of documents containing the search term "Violence", the first order term "Adolescent" as well as the second order term "Right-wing radicalism" is 405.

The goal of the heat map visualization is to enable the user to indicate mainstream areas of the research field more easily. This is relevant for the case where a user starts with a search term and needs an overview of main issues of the field in question. Figure 2 displays, for the example from Figure 1, a heat map of indexing terms that are closely related to the search term "Violence". The column headings of the heat map show the 10 first order terms that most frequently co-occur with the search term. The terms are displayed in descending order of the frequency of their co-occurrence with the search term, i.e. "Adolescent", followed by "Developing country", "Propensity to violence" and so on. The row headings of the heat map display the top three second order terms, i.e. the three indexing terms that most frequently co-occur with the respective first order term as well as with the search term[6]. By this, the heat map visualizes tuples of indexing terms as well as their "heat" with respect to the search term.

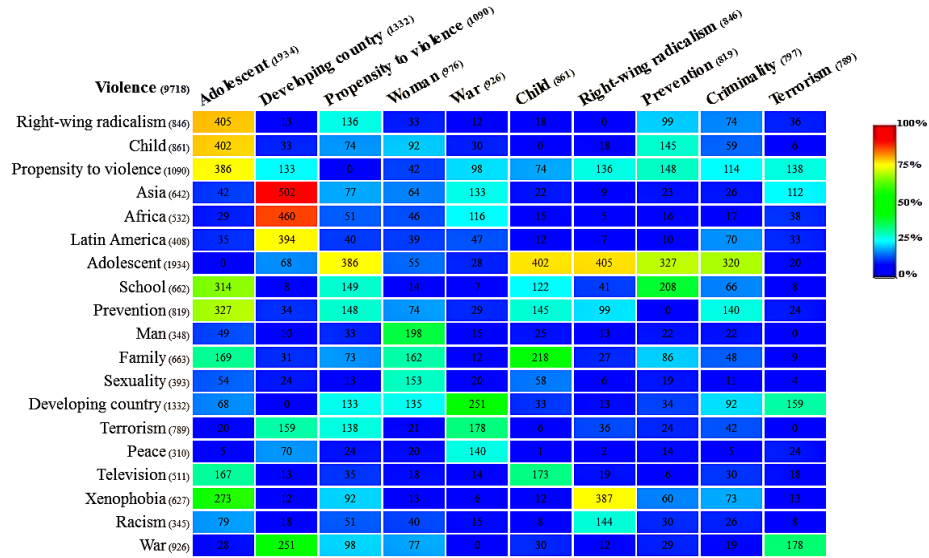

**Fig. 2.** Heat Map Visualization of Co-Word Relationships. The map is divided into "hot" (red), "warm" (yellow, green) and "cold" (blue) areas according to the frequency of co-occurrences. Red cells point to topic combinations which appear most frequently, blue cells correspond to topic combinations which appear least frequently in the map.

The heat map displayed in Figure 2 clearly indicates the combination "Developing country"+"Asia" and "Developing country"+"Africa" as "hot" areas of research in the context of „Violence", followed by "Adolescent"+"Right-wing radicalism", "Adolescent"+"Child", "Developing country"+"Latin America", "Right-wing radicalism"+"Xenophobia" and "Propensity to violence"+"Adolescent". By clicking a term or into a cell the user can browse the documents related to topic combination in question.

---

[6] The list of second order terms is reduced to a disjoint set. For this example, this yields a list of 19 terms (instead of 10x3=30 terms).

It is important to note that instead of visualizing the entire information space by a heat map, the approach here is to reduce the information space to the fraction that matches the information need of the user. The basic idea is to dynamically adapt the heat map of term recommendations when the user clicks into the map or changes search terms, i.e. throughout the entire retrieval process.

## 3      Conclusion and Future Work

The paper proposes a use case for using heat map visualizations of term co-occurrence matrices which can be used as a visual navigation tool through an information space. The approach is to provide a "big picture" view of the relevant fraction of the information space that can be adapted dynamically during a search session. Contrasting areas of different "heat" within a set of co-word relationships highly associated with the user's search term enables the user to indicate main issues of the field in question more easily. We assume that this approach might also help the user to better locate a particular information need within a larger information space. The paper may also provide a principle idea of how knowledge maps of information spaces can be integrated in information seeking processes.

For future work we intend to evaluate the proposed approach with real users on the basis of an evaluation panel [20], apply the approach also to result sets, find more suitable metrics and color codes for calculating and displaying heat maps, and – finally – develop a generic model for the use of heat map visualizations of term suggestions in interactive retrieval systems.